\documentclass[pra,aps,twocolumn,superscriptaddress,showpacs]{revtex4}
\usepackage{graphicx}
\usepackage{amsmath}

\newcommand{\eq}{\begin{equation}}
\newcommand{\fine}{\end{equation}}

\begin{document}

\title{  \bf \Large  Separability inequalities on N-qudit correlations exponentially stronger than 
local reality inequalities}

\date{Jan. 30,2020}

\author{S. M. Roy}
\email{smroy@hbcse.tifr.res.in} \affiliation{HBCSE,Tata Institute of Fundamental Research, Mumbai}

{\begin{abstract}

Abstract. 
   I derive separability inequalities for Bell correlations of observables in  arbitrary pure or mixed $N$ Qudit states in  
   $D^N$-dimensional state space. I find states (a continuum of states if $D>3$) including maximally entangled states  which 
   violate these inequalities by a factor $2^{N-1}$ ; local reality Bell inequalities are much weaker, their maximum violation 
   being by a factor $2^{(N-1)/2}$.  The separability inequalities allow tests of entanglement of unknown states 
   using only the  measured correlations . 
 
\end{abstract}

 
\maketitle

{\bf Introduction}.  Bell inequalities on correlations of observables measured at two sites with two \cite{Bell1964} or 
more settings \cite{Roy-Singh1978} at each site arise from a ``local hidden variable'' (LHV) representation 
derived from the classical idea of Einstein-Bell local reality  ; the representation and hence the 
inequalities also hold in quantum mechanics for separable states.For a {\bf pure} non-separable or entangled  state  one can 
find observables with correlations violating LHV 
inequalities \cite{Gisin1991}.$N$-qubit states can violate the LHV Bell inequalities by a factor upto $2^{(N-1)/2}$ 
\cite{mermin1990},\cite{roy-singh1991},\cite{ardehalli1992} , and this is the maximum possible violation \cite{Cirel'son}.
However, there exist {\bf separability} Bell inequalities for $N$-qubit states $\rho$ \cite{Separability_S.M.Roy2005} which are much stronger ,e.g.
\begin{equation}
 If\>\rho\>separable,\>|Tr \rho \bigotimes _{r=1}^N (\sigma_x -i \sigma_y)^{(r)}|\leq 1\>,\label{NqubitBell}
\end{equation}
and other equivalent inequalities of this form with the Pauli matrix combination $(\sigma_x -i \sigma_y)^{(r)}$ for the $r$-th qubit replaced by  
any unitary transform $U(r)(\sigma_x -i \sigma_y)^{(r)}U(r)^\dagger$. These
inequalities can be violated by a factor upto $2^{N-1}$. 

The extension from qubits to qudits is not easy. 
The proof that LHV Bell inequalities for two qudits  can be violated  \cite{Mermin_Spins}, by a factor $\sqrt{2}$ \cite{Gisin-Peres-Spins}  and 
for $N$ qudits by a factor $2^{(N-1)/2}$ \cite{Cabello-GHZ} involved very ingenious choice of observables. The purpose of 
the present work is to derive a set of {\bf separability} Bell inequalities for arbitrary pure or mixed states of $N$-qudits 
of arbitrary dimension $D$,generalizing Eq.(\ref{NqubitBell}) and to  exhibit quantum states which violate them by a factor $2^{N-1}$.
The violation is an exponentially stronger indication of 
entanglement than the LHV violations for large $N$.  The separability Bell inequalities only involve experimental correlations 
and  are therefore useful to detect entanglement in an unknown quantum state even when  theoretical knowledge ,
e.g. of eigen values of matrices related to the density matrix are not available. 

A {\bf mixed} entangled state need not necessarily violate any LHV inequalities \cite{Werner1989}.
There are many other signatures or measures of entanglement \cite{Horodecki_review_entanglement} such as violation of positivity of 
the partial transpose (PPT) of the density matrix \cite{Peres-partial-transpose}, von Neumann entropy of sub-system density matrices 
\cite{entropy based entanglement}, concurrence \cite{concurrence}  ,  negativity \cite{negativity}, and some criteria based on 
statistical correlation coefficients \cite{Pearson}.In a seminal paper 
Peres \cite{Peres-partial-transpose} made the extremely useful and simple observation that the partial transpose of a separable density matrix 
must be positive. E.g. consider a mixture of a maximally entangled state and the maximally mixed state, the 
$N$- qudit Werner state $\rho_W^{(N,D)}$ where $D$ is the qudit dimensionality,
\begin{eqnarray}
 \rho_W^{(N,D)} =&(p/D)&\sum_{i=0}^{D-1} |i,i,..,i>\sum_{l=0}^{D-1}<l,l,..,l| \nonumber\\
 &+&\frac{1-p}{D^N} {\bf 1}\>,\>0<p<1\>,\label{W(ND)}
\end{eqnarray} 
with {\bf 1}  denoting the $D^N$-dimensional unit matrix. 
Here $|i,i,..,i>\equiv |i>|i>..|i>$ is an $N$-qudit state, and $|0>,|1>,..,|D-1>$ are the $D$ orthonormal states of a single qudit.
E.g. these could correspond to the $2J+1=D$ eigenstates with $J_z=m$ for a particle of spin $J$ ,
\begin{equation}
 |i>=|J,m=J-i>\>.
\end{equation}
 Using PPT Peres proved that the two qubit Werner state  ( $N=2,D=2$ ) is not separable if  $p>1/3$, whereas the Bell-CHSH LHV inequalities 
are violated in the smaller region $p> 1/\sqrt{2}$ \cite{HorodeckiBell} .
Further, the necessary conditions for separability given by PPT are also sufficient for $2\times 2$ and $2\times 3$ dimensional 
Hilbert spaces \cite{HorodeckiPPT} but {\bf not} for higher dimensions or higher number of parties.  

For general $N$ and $D=2$, expansion in overcomplete sets was used by Braunstein et al \cite{WernerNqubit} 
to prove that,
\begin{eqnarray}
\rho_W^{(N,2)}\>is\> separable\> &if&\> p<\frac{1}{1+2^{2N-1}}\>,\nonumber\\
 and\>non-separable\> &if&\> p>\frac{1}{1+2^{N/2}}\>.\label{non-separableN2}
\end{eqnarray}
Intensive study of qudit entanglement \cite{Qudits} using PPT and related Cauchy-Schwarz inequalities \cite{WernerND} 
yield the necessary and sufficient condition for the Werner state to be non-separable for general $D$,
\begin{equation}
 \rho_W^{(N,D)}\> non-separable \>\> iff\>  \>p>\frac{1}{1+D^{N-1}}\>.\label{Best_Werner}
\end{equation}
 For $N>2$, the result(\ref{non-separableN2}) can be substantially improved by using the separability Bell inequality (\ref{NqubitBell}) 
 which implies that 
 $$\rho_W^{(N,2)} \>is\> \>non-separable\> if\>p>2^{-(N-1)},$$
 which is only slightly weaker than the best possible result (\ref{Best_Werner}).
 This , (plus the fact that the `best' results need detailed additional knowledge such as in  PPT tests), motivate us to seek 
 a separability Bell inequality stronger than the local hidden variable inequality for any unknown state of $N$-Qudits.

 {\bf Separability inequalities on Bell-type correlations of $N$ qudits }. 
 Cabello \cite{Cabello-GHZ} constructed the elegant Cabello-GHZ states, and `maximal observables' to 
show maximal LHV violation for $N$-qudits. Instead I  need to introduce   
 a large class of observables which allow strong inequalities on their expectation values in any separable 
 state of odd or  even number of qudits.
 
   If $D$ is even , the basis states $|i>$ of each qudit may be divided into 
  $D/2$ non-overlapping pairs ,each such pairing being labelled by $I$ ,
 \begin{equation}
 I=\{(i_1,j_1),...,(i_{D/2},j_{D/2})\},\>\> i_r<j_r\>; \>D\> even \>,
 \end{equation}
e.g. $I=(0,1),(2,3),..(D-2,D-1) $ .The total number $N_I$ of choices of $I$ is the number of permutations 
of the $D$ indices ,except those which permute the $D/2$ pairs, or 
 the two indices of each pair ,
 \begin{equation}
  N_I= \frac{D!}{2^{D/2}(D/2)!}=1.3.5...(D-1)=(D-1)!!
 \end{equation}
If $D$ is odd, there will be one unpaired index $k$, and we may define , 
\begin{equation}
I=\{k,(i_1,j_1),..,(i_{(D-1)/2},j_{(D-1)/2})\}\>, i_r<j_r\>,
\end{equation}
 Since the unpaired $k$ can be chosen in $D$ ways, the total numbers of choices of $I$ is, 
 \begin{equation}
  N_I=(D-2)!! D\>,\>if\>D\>odd.
 \end{equation}
For the above choices of $I$, I define the operators 
\begin{eqnarray}
&&   \sigma_I=2\sum_{r=1}^M|i_r><j_r| + \eta |k><k|\>,i_r<j_r\>,\label{sigma_I} \\
&&M=D/2\>,\eta=0\>if\>D\>even\>;\nonumber\\
&&M=(D-1)/2\>,|\eta|=1\>,if\>D\>odd\>;\label{M}
\end{eqnarray}
 With these definitions of $M,\eta$, pure qudit states for $D$ even, and odd are of the form,
\begin{equation}
|\psi> =\sum_{r=1}^M \big(c_{i_r}|i_r> +c_{j_r}|j_r>\big)\>+\eta c_k|k>\>,
\end{equation}
where, the normalization conditions are,
\begin{equation}
 |\eta| |c_k|^2+\sum_{r=1}^M \big(|c_{i_r}|^2 +|c_{j_r}|^2\big)=1\>,
\end{equation}
 This yields the basic inequalities,
\begin{eqnarray}
&& |<\psi|\sigma_I|\psi>|= |\>\eta |c_k|^2+\sum_{r=1}^M 2 c_{i_r}^*c_{j_r} |\nonumber\\
&& \leq |\eta||c_k|^2+\sum_{r=1}^M |c_{i_r}|^2 +|c_{j_r}|^2=1\>,\label{sigmaIineq}
 \end{eqnarray}
 It will be useful to introduce Hermitian operators $\sigma_I^\pm$ which obey, 
 \begin{eqnarray}  
&&\sigma_I^+  \equiv \frac{\sigma_I +\sigma_I ^\dagger}{2},\>\sigma_I^-\equiv \frac{\sigma_I - \sigma_I ^\dagger}{2i} ,\\
 && (\sigma_I^+)^2={\bf 1}-(Im \eta)^2|k><k|\>.\label{sigmasqI+}\\
 &&(\sigma_I^-)^2={\bf 1}-(Re \eta)^2|k><k|\>.\label{sigmasqI-}
 \end{eqnarray}
Since $|\eta |\leq 1$, equations (\ref{sigmasqI+}),(\ref{sigmasqI-}) show that $\sigma_I^\pm $ have eigen values $\in (-1,1)$ . 
 
 Consider now a separable state $\rho$ of $N$ qudits; it can always be written as 
 a convex sum of factorizable pure states ,
 \begin{equation}\label{separable}
  \rho = \sum_s p_s \bigotimes_{n=1}^N |\psi_s^{(n)}><\psi_s^{(n)}|\>,p_s\> >0,\>\sum_s p_s=1\>. 
\end{equation}
In analogy to Eqs. (\ref{sigma_I}),(\ref{M}) I define the operators $\sigma_{I_n}^{(n)} $ for the $n$-th qudit , 
where 
the index sets $I_n$ for each of the qudits can be chosen independently. 
\begin{eqnarray}
  &&\sigma_{I_n}^{(n)}=2\sum_{r=1}^M(|i_r><j_r|)^{(n)}\nonumber\\
 && +(\eta |k><k|)^{(n)} \>,(i_r)^{(n)}<(j_r)^{(n)}\>,\label{sigma_I_n}
    \end{eqnarray}
 Now define the $N$-qudit operators and their expectation values in the separable state (\ref{separable}),
 \begin{eqnarray}
 &&\Sigma_I \equiv \bigotimes_{n=1}^N \sigma_{I_n}^{(n)}=\bigotimes_{n=1}^N \{(\sigma_{I_n}^+)^{(n)}+i(\sigma_{I_n}^-)^{(n)}\},\label{SigmaI}\\
  && Tr \rho \>\Sigma_I\> = \sum_s p_s \prod _{n=1}^N <\psi_s^{(n)}|\sigma_{I_n}^{(n)}|\psi_s^{(n)}>.
\end{eqnarray}
The basic inequalities (\ref{sigmaIineq}) applied to each of the qudits, and the conditions $p_s>0,\sum_s p_s=1 $ now yield, for 
every separable $N$-qudit state,
\begin{equation} \label{BellI}
 |Tr \rho \>\Sigma_I\>|\leq 1\>;  
\end{equation}
or in terms of Hermitian operators, 
\begin{eqnarray}  
&&\Sigma_I^+  \equiv (\Sigma_I + \Sigma_I ^\dagger)/2,\> \Sigma_I^- \equiv (\Sigma_I - \Sigma_I ^\dagger)/(2i),\\
 &&|Tr \rho \>\Sigma_I^+\>|^2+|Tr \rho \>\Sigma_I^-\>|^2\leq 1\>,\label{BellI'}
 \end{eqnarray}
This is a large set of  necessary conditions for separability, a total of $\big((D-1)!!)\big)^N$ for even $D$,
and $\big((D-2)!! D\big)^N$ for odd $D$, because the index sets  
$I_n^{(n )}$ 
for each of the $N$ qudits can be chosen independently.
 From the Hermitian and antihermitian parts of (\ref{SigmaI}) we see that $\Sigma_I^\pm$  are linear 
 combinations of $2^{N-1}$ terms, 
each term being a product of the commuting Hermitian operators with eigen value $\in (-1,1)$  for the $N$ different qudits.
Hence Eqns.  (\ref{BellI}) or(\ref{BellI'}) are quadratic constraints on $N$-qudit Bell-type correlations in separable states $\rho$. 
They imply the weaker linear inequalities,
\begin{eqnarray}
&&|Tr \rho \>\Sigma_I^+\>|\leq 1\>;|Tr \rho \>\Sigma_I^-\>|\leq 1\>;\nonumber\\
&&|Tr \rho \>\Sigma_I^+\>|+|Tr \rho \>\Sigma_I^-\>|\leq \sqrt{2}.\label{linear Bell}
\end{eqnarray}
I shall find a large number of states (a continuum of states if $D>3$ ) which violate the LHV inequalities maximally, i.e. 
by a factor $2^{(N-1)/2}$, and violate the separability inequalities by a factor $2^{N-1}$.

 {\bf Violation of separability inequalities and LHV inequalities by entangled $N$-qudit states}.
  The LHV representations for the expectation values of the observables (\ref{SigmaI})  are,
  \begin{eqnarray}
 && Re (\Sigma_I)_{LHV}+i Im (\Sigma_I)_{LHV}=\int d\lambda \rho (\lambda) \nonumber\\
&&\prod _{n=1}^N \{(\sigma_{I_n}^+)^{(n)}(\lambda)+i(\sigma_{I_n}^-)^{(n)}(\lambda)\},
    \end{eqnarray}
    where $(\sigma_{I_n}^\pm)^{(n)}(\lambda) \in(-1,1)$, and $\rho (\lambda)$ is 
    a normalized probability density.By varying the $(\sigma_{I_n}^\pm)^{(n)}(\lambda)$ in the allowed range 
    we obtain the LHV inequalities \cite{mermin1990},\cite{roy-singh1991},\cite{ardehalli1992} ,
    \begin{eqnarray}
 N odd\>&:&  |Re (\Sigma_I)_{LHV}| \leq \sqrt{2}^{N-1}\>;\label{LHV1}\\ 
\>\>\>\>\>\>&&|Im (\Sigma_I)_{LHV}| \leq \sqrt{2}^{N-1}\>;\label{LHV2}\\
 N even &:&|Re (\Sigma_I)_{LHV}| + |Im (\Sigma_I)_{LHV}| \leq \sqrt{2}^{N}\label{LHV3}
\end{eqnarray} 
I now construct a continuum of $N$-qudit states violating these inequalities maximally. 
I start from the normalized $n$-th qudit states $|+>^{(n)},|->^{(n)}$ ,which are 
arbitrary superpositions of mutually orthogonal sets of states $\{(|i_r>)^{(n)} \},\{(|j_r>)^{(n)} \}$ introduced 
before, each depending on $M$ complex parameters $\{(c_{i_r})^{(n)} \}$ ,

   \begin{eqnarray}
&& |+>^{(n)}=\sum_{r=1}^M (c_{i_r})^{(n)}(|i_r>)^{(n)}\>, \\
&&|->^{(n)}=\sum_{r=1}^M (c_{i_r})^{(n)} (|j_r>)^{(n)},\\
&&\sum_{r=1}^M |(c_{i_r})^{(n)}|^2=1\>.\label{normalization}
\end{eqnarray}
 The definition (\ref{sigma_I_n}) yields,
 \begin{eqnarray}
&&(\sigma_{I_n})^{(n)}|->^{(n)}=2 |+>^{(n)}\>;\\
&&\big((\sigma_{I_n})^{(n)}\big)^\dagger|+>^{(n)}=2 |->^{(n)}\>.
\end{eqnarray}
I now define the $N$-qudit states,
\begin{eqnarray}
&&|\pm,..,\pm>\equiv\bigotimes_{n=1}^N |\pm>^{(n)},\\
 && |\psi(\mu) >=\frac{|+,..,+> +\mu|-,..,->} {\sqrt{2}}\>,|\mu|=1\>.\label{psimu}
 \end{eqnarray}
 Using,
 \begin{eqnarray}
  &&\Sigma_I |-,..,->=2^N|+,..,+>\>;\\
&&(\Sigma_I)^\dagger |+,..,+>=2^N|-,..,->\>,
 \end{eqnarray}
and setting successively $\mu=\pm 1,\pm i,e^{\pm i\pi/4} $ , I obtain the eigen value equations,
 \begin{eqnarray}
  \Sigma_I^+ |\psi(\pm 1) >=\pm 2^{N-1}|\psi(\pm 1) >\>;\label{QM1}\\
  \Sigma_I^-|\psi(\pm i) >=\pm 2^{N-1}|\psi(\pm i) >\>,\label{QM2}\\
  (\Sigma_I^+ \pm \Sigma_I^-)|\psi( e^{\pm i\pi/4}) >=2^{N-1/2}|\psi( e^{\pm i\pi/4}) >.\label{QM3}
    \end{eqnarray}
 
Eqns. (\ref{QM1}),(\ref{QM2}),(\ref{QM3}) demonstrate  quantum violations of the LHV predictions (\ref{LHV1}),(\ref{LHV2}),(\ref{LHV3}) 
respectively by the maximum possible factor $\sqrt{2}^{N-1}$.

The same states yield also,
\begin{equation}
 <\psi(\mu)|\Sigma_I|\psi(\mu)>=\mu 2^{N-1}\>, |\mu|=1,
\end{equation}
demonstrating violations of the separability 
inequality (\ref{BellI}) by a factor $2^{N-1}$; further, Eqns. (\ref{QM1}),(\ref{QM2}),(\ref{QM3}) 
show violations of the weaker linear separability inequalities (\ref{linear Bell}) by a factor $2^{N-1}$ which is exponentially larger than 
the maximum possible violation of the LHV inequalities. 
 
 {\bf Violation of separability Bell inequalities by maximally entangled states and Werner states}.
 
In our notation, the maximally entangled $N$-qudit state  \cite{Qudits},\cite{WernerND} has the representation,
  \begin{eqnarray} 
  && |\psi_{max}>=\frac{1}{\sqrt{D}}\big(\zeta |k,k,..,k>\nonumber\\
  && + \sum_{r=1}^M (|i_r,i_r,..,i_r> +|j_r,j_r,..,j_r>)\big),
  \end{eqnarray}
where $\zeta=0$ for $D$ even, and $\zeta=1$ for $D$ odd. I Choose the definition (\ref{sigma_I_n})  with 
$|i_r>^{(n)}=|i_r>,|j_r>^{(n)}=|j_r>$, and $\eta=1$ if $D$ odd, $\eta=0$ if $D$ even. Defining 
$\Sigma_I$ as in Eqn. (\ref{SigmaI}) we have,
\begin{eqnarray}
   &&D\> odd\>\>:\> <\psi_{max}|\Sigma_I |\psi_{max}>= 2^{N-1}\>\frac{D-1}{D} +\frac{1}{D},\nonumber\\
   &&D\> even\>:\> <\psi_{max}|\Sigma_I |\psi_{max}>= 2^{N-1}\>.
\end{eqnarray}
Hence, for even $D$, the separability inequality (\ref{BellI}) is violated by the maximally entangled state by
a factor $2^{N-1}$, and for odd $D$ by a slightly smaller factor.

For the Werner state (\ref{W(ND)}) we find that,
\begin{eqnarray}
 &&Tr \rho_W^{(N,D)} \Sigma_I =p\big(2^{N-1}\frac{D-1}{D} +\frac{1}{D}\big)\nonumber\\
 &&+\frac{1-p}{D^N}\>;for\>D\>odd\>;\\
 &&Tr \rho_W^{(N,D)} \Sigma_I =p\>2^{N-1}\>;for\>D\>even\>.
\end{eqnarray}
Thus the separability Bell inequality (\ref{BellI}) is violated by the Werner state (\ref{W(ND)}) 
 \begin{eqnarray}
 && \>if\>p >2^{-(N-1)}, \>for \>D\> even,\\
 &&\> if \>p>\frac{D^N-1}{(2D)^{N-1}(D-1)+D^{N-1}-1}\>,D\>odd\>,
 \end{eqnarray}
 which ,in addition to validity for arbitrary $D$ constitute an exponential improvement over the  
   $N$ qubits result (\ref{non-separableN2}) of Braunstein et al \cite{WernerNqubit} for large $N$. 
   This result is however weaker than the best possible result (\ref{Best_Werner}) which uses detailed 
   theoretical knowledge such as PPT tests on the density matrix.
   
   {\bf Conclusion}. I have obtained necessary conditions for separability of unknown $N$-qudit states in the form of 
   inequalities on Bell correlations (\ref{BellI}),(\ref{BellI'}); there are a total of $\big((D-1)!!)\big)^N$ 
   inequalities for even $D$, and $\big((D-2)!! D\big)^N$ inequalities for odd $D$. The inequalities can be violated by 
   a factor $2^{N-1}$ by quantum states, whereas the maximum possible violation of LHV Bell inequalities is by a 
   factor $2^{(N-1)/2}$. A novel feature for qudits with $D>3$ is that the quantum states showing violation of 
   the separability Bell inequalities by a factor $2^{N-1}$ form a continuum of states depending on $M^N$ complex parameters 
   constrained only by $N$ normalization conditions (\ref{normalization}), where $M=D/2$ for even $D$, and $M=(D-1)/2$ for odd $D$.
   The maximally entangled state ,viz. the state (\ref{W(ND)}) with $p=1$ is shown to violate 
   the seaparability inequalities (\ref{BellI}) by a factor $2^{N-1}$ for even $D$, and by a slightly smaller factor for odd $D$.
   The Werner state (\ref{W(ND)}) is shown to violate the separability Bell inequalities if $p>2^{-(N-1)}$ if $D$ is even.
   Testing the separability Bell inequalities, like the LHV Bell inequalities and unlike many other criteria, need knowledge of only 
   the experimental Bell correlations, and no other theoretical knowledge about the state. The separability Bell inequalities 
   are however exponentially stronger than the LHV inequalities. Are there observables different from the $\Sigma_I$, eqn. (\ref{SigmaI}), 
   which lead to stronger separability inequalities ? This question is left open.
   
   Acknowledgement. I thank Ritwick Ghosh and Ajinkya Werulkar for preliminary discussions on this work.
   I thank the Indian National Science Academy for the INSA honorary scientist position at the Homi 
   Bhabha Centre for Science Education, TIFR.

\end{document}